\documentclass{article}

\begin{document}

\title{Gauge and General Relativity}

\author{M.D. Iftime \thanks{MCPHS and Boston U., Massachusetts, USA}}

\date{}
\maketitle

\begin{abstract}

One of the main features of covariant theories, in particular general relativity, is that the field equation possesses gauge freedom associated
with global diffeomorphisms of the underlying manifold. I shall explain here how the hole argument is a reflection of this gauge freedom.

Finally I shall point out some implications of the hole argument and extend its formulation to the case of permutable theories. 

As covariant theories provides a general mathematical framework for classical physics, permutable theories provide the language for quantum physics. Permutable theories are defined functorially on the category of sets and permutations with values into the category of fibered sets and fiber-preserving automorphisms, and rules of selecting sections. The hole argument for permutable theories is intimately related with the individuation problem of the base elements ( e.g., elementary particles). This is a consequence of the fact that the automorphisms of the base space provides the gauge freedom of the fields. 

\end{abstract} 


\pagebreak

\tableofcontents

\pagebreak

\section{Introduction}

Fiber bundles provide an adequate geometrical formulation for classical field theory, where fields on a smooth spacetime \footnote{We call $M$ a spacetime manifold, however we did not assume any fix spacetime metric given a priori on $M$} manifold $M$ are represented by sections of some fiber bundle $(E\to M)$ over $M$. The configuration space $E$ of the bundle is selected by the field equations, which are required to determine the configuration of the physical system for given initial conditions. We wish to analyse the case when physical systems admit gauge transformations. To illustrate this, we shall start with two familiar examples:

\subsection{Gauge Theories}

All classical fundamental theories have a large symmetry group. This group results in underdetermination and/or overdetermination of the field equations which is in turn deeply related to what can be physically observed or equivalently, what is a genuine property of the physical world and what is just a convention of the observer.

Yang Mills theory provide a universal description of the fundamental electroweak and strong interactions. It is a gauge theory of principal
connections on a principal bundle $(P\to M; G)$. Being $G$-equivariant,
principal connections on $P$ i.e., gauge potentials, are identified to global sections of the gauge natural bundle (e.g., \cite{KMS},
\cite{Fatibene}) $E = J^{1}P/G$, where $J^{1}P$ denotes  the first order jet manifold of sections of $(P\to M)$. The role played by the structure bundle $(P\to M; G)$ is that it selects a class of global sections.
Gauge transformations in Yang Mills gauge theory are vertical automorphisms
of the principal bundle that project over the identity base diffeomorphism
$id_{M}:M\to M$. They are in 1:1 correspondence with global sections of the group scheme $(AdP\to M)$, the principal bundle associated with the adjoint representation of $G$ on itself. The (pure) gauge group $Gau(P)$ acts on associated bundles: each gauge transformation $(\phi_{P}, id_M)$ induces an automorphism $(\phi_{E}, id_M)$, $\phi_{E}:(P\times F)/G\to(\phi_{P}(P)\times F)/G$ of the associated bundle, but $Diff M$ does not. The reason for this is that there may be diffeomorphisms of $M$ that are not projections of a principle automorphism, i.e., there is no
map\footnote{Except in the case of principal frame bundles.} from $Diff M$ to $Aut(P)$. \footnote{$Aut(P)$ denotes the group of general principal automorphisms of $(P\to M)$. There is an exact sequence of group homomorphisms $\{e\}\to Gau(P) \stackrel{i}{\hookrightarrow} Aut(P) \stackrel{pr}{\longrightarrow } Diff M\to \{e\}$.}.

Gauge morphisms transform sections into sections, and solutions into solutions ( i.e., they are symmetries of the field equation. The Lagrangian of the gauge theory is invariant under the action of the gauge group $Gau(P)$)

The field equations are required to determine the configuration of the physical system for given initial conditions. However because of the gauge freedom, one produces a whole gauge-related family of solutions, with the same initial conditions, instead of a unique solution.

{\em The Hole Argument for Gauge Theories:}

Solving locally the field equations one can produce two distinct solutions that are not required to coincide on the overlapping domain, but they must agree up to a gauge transformation, so that they can still represent an unique physical object. In this way, one can produce a whole family of gauge-related different solutions on $M$ with the same initial conditions. 

In conclusion, classical gauge theories are undetermined: gauge theories can only determine a class of gauge-related solutions, not a representative of the class (unique solution).  To avoid this version of the hole argument, one must assume that two different mathematical models (solutions) which differ by a (compact supported) gauge transformation represent the same physical object.

If we denote by $\mathcal{M}_{G}$ the moduli space of solutions of a $G$(gauge) natural bundle (representing the configuration space of a gauge theory), the gauge group $Gau(P)$ acts as a transformation group on $\mathcal{M}_{G}$, partitioning $\mathcal{M}_{G}$  into (disjoint) gauge-related classes of solutions.\footnote{The gauge group $Gau(P)$ is isomorphic with the group of global sections $\Gamma(M; AdP)$ of the group scheme $AdP$. In some cases, one can define Sobolev completion of the group $Gau(P)$ to a Lie group.} A physical gauge field represents a point in the orbit space $\mathcal{M}_{G}/Gau(P)$, so that the canonical projection $\Pi:\mathcal{M}_{G}\longrightarrow \mathcal{M}_{G}/Gau(P)$ identifies all gauge-related solutions to a representative of that class. The image of $\Pi(\sigma)=[\sigma]$ represents the (unique) physical gauge field defined by $\sigma$.

\subsection{General Relativity}

General Relativity and, in general any covariant theory\footnote{Some authors define general covariance as what we call covariance, not making a clear distinction between the two.\cite{Stachel1993}, 
\cite{Iftime_Stachel2006}}, admit covariant transformations. Fiber bundles possessing such transformations belong to the category of natural bundles\cite{Nijenhuis}. Given a natural bundle $(E\to M)$ over a base manifold $M$, there exists a canonical lift of any global/local diffeomorphism of $M$ to a bundle automorphism of $E$ called a covariant/natural transformation. The group $Diff M$ acts on the bundle $(E\to M)$ taking sections into sections, and solutions into solutions (i.e., they are symmetries of the field equations; The Langrangian of the system is invariant under the action of $Diff M$).
This implies that the gravitational field equation possesses gauge freedom associated with global diffeomorphisms of the underlying manifold. As for gauge theories, a consequence of this gauge freedom of Einstein theory, the gravitational field equation does not admit an initial-value formulation in traditional sense, but only up to diffeomorphisms. This shows the intimate relationship between the "gauge freedom" of Einstein's theory and the hole argument. 

{\em The Original Hole Argument:}

Einstein (1913) formulated the hole argument in coordinate language as a boundary-value problem. Hilbert (1917) reformulated the hole argument an initial value problem. The two formulations are known as the boundary value and initial value formulations of the original hole argument.
\cite{Stachel2001} The two formulations of the hole argument translated in the manifold language are:

{\em (1) The boundary-value formulation of the hole argument:}

Let $(M,g)$ be a spacetime manifold containing a "hole",  i.e., bounded, closed spacetime region $H$ on which the metric field $g$ is the only one present, so that inside $H$, the metric $g$ obeys the (homogeneous) Einstein's empty field equations:

\begin{equation}
Ein(g)=Ric(g) -\frac{1}{2}gR(g)=0
\end{equation}

Given a solution $g(x)$ everywhere outside of and on the boundary of $H$, including all the normal derivatives of the metric up to any finite order on that boundary, this data still does not determine a unique solution inside $H$ (no matter how small $H$), because an unlimited number of other solutions can be generated from it by those diffeomorphisms that are identity outside $H$ (and any number of derivatives of which also reduce to the identity on the boundary), but differ from the identity inside $H$. The resulting metric $g'(x)$ will agree with $g(x)$ outside of and on the boundary of $H$, but will differ from it inside $H$.\\

{\em (2) The initial-value formulations of the hole argument:}\\
Assume that initially the universe was filled with matter and that later in time a hole forms (i.e., a region without matter). Let $g(x)$ and $g'(x)$ be two distinct solutions of the field equations which are equal everywhere in $M$ except for a hole, then if assume a spacelike (initial data) surface $S_{3}: t=0$ such that the hole is entirely in the future development $D^{+}(S_{3})$ of it, then because the two metrics are equal everywhere outside and on the boundary of the hole, they will have the same set of initial data on the surface. \cite{Rovelli1999}

In conclusion, there is no well-posed initial-value and/or boundary-value problem for Einstein's covariant equations. General reltivity cannot determine the space-times metric field! It seems that we uncovered a mystery of the theory!

Einstein found these results unacceptable and claimed in 1912 that {\em general covariance} cannot be a property of the theory of gravity! He spent the next three years looking for non-generally covariant field equations.

To solve the puzzle of his hole argument, Einstein applied Dirac's postulate\footnote{only gauge invariant( or Dirac observables) can be measurable quantities} to his field equations. His way out of the hole argument is to conclude that two such space-time metrics $g(x)$ and $g'(x)$ represent the same gravitational field  or in other words, general relativity is a generally-covariant theory, but there is a mistaken assumption about the nature of space and time. After dropping that assumption there will no longer be any incompatibility between general covariance of his field equations and determinacy.

In the language of manifolds, Einstein's line of reasoning on how to avoid the hole argument is to assume that at least inside the hole, the space-time points are not individuated independently of the metric field. This assumption automatically implies general covariance of the field equations, which translates in diffeomorphism invariance.

Indeed, if the points of the space-time manifold $M$ are not individuated independently of the metric, it implies that

\begin{quote}

"..when we drag-along the solution, we actually drag-along the physically individuating properties and relations of the points. In other words, the space-time points have no inherent chrono-geometrical or inertio-gravitational properties or relations that don't depend on the presence of the metric." 

(John Stachel)
\end{quote}

So, the pull-back metric does not differ {\em physically} from the original one, and an entire equivalence class of diffeomorphically-related solutions to the field equations should corresponds to {\em one} inertio-gravitational field.

As a consequence, 

\begin{quote}

...the points of the manifold can be characterized as space-time elements, but they lack individuation as events, i.e., points of a particular space-time, unless and until the metric field is specified. (John Stachel)
\end{quote}

Mathematically, if we denote by $\mathcal{M}(M)$ the collection of spacetime metrics on $M$. The group $Diff(M)$ acts as a transformation group on $\mathcal{M}(M)$ by pulling back metrics on $M$: for all $\phi\in Diff(M)$ and $g\in \mathcal{M}(M)$ the action map is defined by $(\phi, g)\longmapsto \phi^{*}(g) $. \footnote{$Diff(M)$ is a smooth manifold modelled on the space of vector fields on $M$, and acts naturally on all tensor bundles over $M$ by differentiation, and so on any sections of these bundles.}
The action of $Diff(M)$ on $\mathcal{M}(M)$ partitions $\mathcal{M}(M)$ into (disjoint) isometry classes of metrics. A hysical gravitational field represents a point in the superspace\footnote{see e.g.\cite{Fisher}} $\mathcal{M}(M)/Diff(M)$. The canonical projection $\Pi: \mathcal{M}(M)\longrightarrow \mathcal{Q}(M)$ identifies all diffeomorphically related metrics to a representative of that class, so that the image of $\Pi(g)=[g]$ represents the (unique) physical gravitational field defined by $g$. \cite{Iftime_Stachel2006}

\subsection{Generalized Gauge}
 
In the real world, all physical systems coexist on the underlying spacetime manifold $M$, and they are normally in interaction with each other. \cite{Geroch}The above two example treated separately (pure) gauge fields and gravity. It is interesting to analyse the situation in which combined gauge and gravity (or any covariant fields) into one dynamical system. Simple  examples are provided by the coupled Einstein-Maxwell or  Einstein-Yang Mills equations. The configuration space of the combined system is a bundle $(E\to M)$ where the fibres include an antisymmetric tensor (for electromagnetism, for example), and a Lorentz metric (for gravity). The field equations are the (inhomogeneous) Einstein equations $Ein(g)=Ric(g) -\frac{1}{2}gR(g)=T(g,\psi)$, together with the set of dynamical equations for the non-gravitational matter and fields( the coupled Einstein-Maxwell or Einstein-Yang Mills equations), so that the coupled fields equations are covariant.\footnote{The spacetime structure and the source fields constitute a dynamical system, the equations of which can only be solved together.}

However because the diffeomorphism group $Diff M$ does not act on the configuration space of the gauge fields, one must restrict to a subcategory of manifolds, so that it makes possible to lift all base morphisms to bundle automorphisms and treat the gauge fields as 
"natural" objects in the combined system.

{\em The Hole Argument for couple Einstein-Yang Mills Equations}

For the set of coupled gravitational and non-gravitational field equations that are covariant, a generalized version of the hole argument applies straightforward.

There is only one way out of this version of the hole argument: one must assume that two diffeomorphically-related solutions of the coupled field equations represents the same set of physical gravitational and non-gravitational fields.

\section{Permutable Theories}

Up to now we did not consider discrete models. The spacetime arena on which all the action took place was a fixed finite dimensional manifold. Can we further generalize the formulation of the hole argument to a more general class of theories to include discrete cases, permutable theories?

Essentially the argument seems to be independent of the differentiability or continuity properties of the fibered manifold of the covariant theory upon which the hole argument is formulated.

In \cite{Iftime_Stachel2006} we used the category language to generalize the Einstein's hole argument to other covariant theories for natural objects. 

As covariant theories provides a general mathematical framework for classical physics, permutable theories can provide the language for quantum physics. This analogy is deeply founded: as covariant theories can be defined functorially on natural bundles, permutable theories can be defined functorially on the category of sets and permutations with values into the category of fibered sets and fiber-preserving automorphisms, and rules of selecting sections.(\cite{Iftime_Stachel2006})

The hole argument for permutable theories is intimately related with the individuation problem of the base elements ( e.g., elementary particles). This is a consequence of the fact that the automorphisms of the base space provides the gauge freedom of the fields.

\subsection{From Differentiable to Discrete}

We will start by giving the basic definition and notations: 

A fibered set is a surjective map $\pi :X\to S$ between two non-empty sets $X$ and $S$.  
A section of this fiber set is a map $\sigma: S\to X$  such that $\pi \circ\sigma =id_S$. 
A bijection $S\stackrel{\bar{\alpha }}\to S$ is sometimes called automorphism (or permutation) of $S$.\footnote{In category language, we say that $\bar{\alpha }$ has and inverse that is both a section and a retraction.} 
The set of all permutations of $S$ forms the permutation group of $S$, $Perm(S)$. 
An object in the category of permutations consists of a set $S$ together with a given permutation $\bar{\alpha }$; and we will denote it by $S^{\bar{\alpha }}$.
\footnote{ The compostion of such maps is straigthforward.\cite{MacLane1997}}
A fiber-preserving automorphism $X^{\alpha}$ of a fibered set $X$ is a nothing but a permutation $\alpha:X\to X$ of $X$ that preserves the fibers of $X\stackrel{\pi }\to S$.
A fiber-preserving automorphism  $X^{\alpha}$ is naturally associated with a base permutation $S^{\bar{\alpha}}$.  

In analogy to the geometric objects, we restrict ourselves to the case in which to each base permutation $S^{\bar{\alpha}}$ there corresponds a  (unique) "natural" lift to a fiber-preserving automorphism $X^{\alpha}$ (i.e., $\bar{\alpha }$ is the projection of $\alpha $).

A subclass of such fibered sets are $G$- sets, where $G$ is a finite group. Let  $X\stackrel{\pi }\to S$ be a fibered set for which all the fibers contain the same number of elements. An action of $G$ on $X$ is defined to be an homomorphism  $(g,x)\in G\times X\longmapsto x\cdot g\in X$ of $G$ into $Perm(X)$ . 
An equivalence relation $\rho $ on $X$ is $G$-invariant (or a congruence
on $X$) if the action of $G$ on $X$ preserves the relation, i.e., 
if $x=y(\mathrm{mod}\rho)$  then $x\cdot g =y\cdot g(\mathrm{mod}\rho)$
for all $g$ in $G$. A $G$- set is determined by the action of a group $G$ of  permutations of $X$, action that commutes with the projection map $\pi$(
the action of $G$ is a fiber automorphism that permutes the fibers of $X$). When $G = Perm(S)$ we recover the abstraction to fibered sets with permutations of a type of geometrical object.\cite{Iftime_Stachel2006} Since in this case there is a unique action of $G$ on the total set $X$, action that commutes with the fibers, then one can define an equivalence relation on $X$ defined by the projection map $\pi$. For example, principal bundles ( where $G$ is a Lie group of diffeomorphisms) and regular coverings (where $G$ is discrete).

The hole argument that applies to covariant theories modelled on a natural bundle can be modified so it applies to discrete models. One can abstract from the topological and differentiable structures by applying a forgetful functor\footnote{By definition, a (covariant) functor $F : D \to C$ of categories $D$ and $C$, is forgetful if $F$ is injective on $D-$ maps. In other words, given some object with structure as input, some or all of the object's structure or properties is 'forgotten' in the output.} to obtain a lifting (covariant) functor from the category of sets with permutations into the category of fibered sets and fiber-preserving automorphisms.\footnote{"Abstraction by deletion is a straightforward process: One carefully omits parts of the data describing the
mathematical concept in question to obtain the more abstract concept." - \cite{MacLane}}. At this level of abstraction, the total space $X$ and the base space $S$ are discrete sets, and sections are maps $\sigma$ from $X$ to $S$ such that $\pi\circ \sigma=id_{S}$. These models can be used to treat quantum mechanical many-particle systems, in particular the case when these particles are all of the same kind, as we shall discuss later. 

Another level of abstraction is obtained by taking a category of sets over $m$-manifolds, which is defined as a category $\mathcal{C}$ with a fixed covariant functor $F$ from the category ${\mathcal Mf_{m}}$ of $m$-manifolds and local diffeomorhisms into $\mathcal{C}$. If $M$ is a $m$ -manifold (object in ${\mathcal Mf_{m}}$), an object $X$ of $\mathcal{C}$ lifts $S$, or is a lifting of $M$, if $F(X) = M$, and similarly for morphisms. A morphism of categories over $m$-manifolds is a covariant functor commuting with the projection to ${\mathcal Mf_{m}}$.
The category $\mathcal{C}$ is nothing but the subcategory of $\mathcal{FM}$by the pullback of a forgetful functor ${\mathcal Mf_{m}}\to{\mathcal Sets}$. 

Examples of "discrete" fields are continuous fields with a discrete spectrum of solutions. They can be viewed as sections of an appropritate category of sets over manifolds. For example, the wave equation in a $1$-dimensional string. If no boundary conditions are imposed, there are a continuous spectrum of solutions: one for every positive real number representing a possible frequency. If both ends of the string are held fixed, there are only a countable infinity of solutions: the fundamental frequency (no nodes), and all the higher harmonics ( with $1, 2,... n,...$ nodes). In general, any eigenvalue problem will have a discrete set of solutions.

\subsection{Physical Models}

Let $S=\{a_{1},a_{2},..., a_{n}\}$ be a non-empty set of $n$ elements. Assume that the elements of $S$ describe particles of the same type (e.g., all are electrons). We wish to construct a fibered set $(E\stackrel{\pi}\to S)$ over $S$ by defining $E$ to be the cartesian product $S\times F$, where  the fiber
$F$ represents the set of all possible physical states or processes of an arbitrary  particle in $S$. The projection map $\pi$ is given by the projection $pr_1$ over $S$. A fibre over an particle $a\in S$ is the of all possible states or processes at $a$. A section of $(E\stackrel{\pi}\to S)$ is given by a map $\sigma :S\to F$ such that $\sigma(a_{i})=(a_{i}, f_{i})$, where $f_i$ represents a choice of a state or process at $a_{i}\in S$ in the fiber $F$.
A  theory on $S$ can be defined as a rule for selecting a class of sections
of $(E\to S)$ ( solutions). 

If the elements of $S$ describe particles of different type (e.g., protons and electrons), denoting with $F_{i}$ the set of elements of $S$ of the same $i$-th kind (i.e., of the same quiddity), we can write $S=\bigcup_{i=1}^{k}{F_{i}}$. The configuration space $E= F_{1}\times\cdots\times F_{k} $ represents the space of all possible states of the elements in $S$.

A set of relations between elements of different kind in $S$ can be interpreted as an $k$-ary relation $R\subseteq F_{1}\times\cdots\times F_{k}$. A section of $(E\stackrel{\pi}\to S)$ is a map $\sigma :S\to E$ such that 
$(f_{1},\cdots f_{n})$, where $f_i$ is in $i$-th the domain of the $k$-ary relation $R$. A section represents a selection of a state in the fiber $F_{1}\times\cdots\times F_{k}$ that involves $a$, for each $a\in S$.
and can also be viewed as a an $k+1$-ary relation, if we take the image $\sigma(S)=R_{\sigma}\subset S\times F$.

Chosing a section $\sigma$ of $(E\to S)$, then by permuting the elements of the base set $S$, one obtains a pulled back section $\phi^{*}(\sigma)$. 
A theory over $S$ is called permutable, if the all pulled back sections of solution are also solutions. In other words a theory is permutable if {$\{R_{\sigma}\}$ is in the selected class of models, so is $P\{R_{\sigma}\}$ for all permutations $P$ of the base set $S$. 

Do the two such equivalent solutions represent the same physically? 
The answer to this question is purely physical. If "yes", then the theory is generally permutable. 

In a generally permutable theory permutations of 'identical' particles are symmetries. They are transformations that leave the physical situation unchanged, so it requires that the relations of our formal language be invariant under permutations. 

{\em The hole argument:}

The hole argument can be straightforward formulated for generally permutable theories. Let consider a permutable theory defined on a non-empty set $S$, i.e. a rule for selecting sections of a fiber set $(E\stackrel{pr_{1}}\to S)$ as above. Based on a similar construction of the quotient space of classes of models of a covariant theory\cite{Iftime_Stachel2006}, there is  a 1:1 correspondence between a point in the quotient space ( i.e., a class of authomorphically related mathematical models) and a (unique) physical model of the theory.

If we denote $\mathcal{M}(S)$ the collection of solutions of a permutable theory. Then the group $Perm(S)$ acts as a transformation group on $\mathcal{M}(S)$ by pulling back solutions $(f, \sigma)\longmapsto f^{*}(\sigma) $ for all $f\in Perm(S)$ and $\sigma\in \mathcal{M}(S)$.

For a fixed section $\sigma$, then    $\mathcal{O}_{\sigma}=\{\phi^{*}(\sigma)|\quad \phi\in Perm(S)\}$ 
is the set of all automorphically-related sections with $\sigma$. The group $Perm(S)$ partitions $\mathcal{M}(S)$ into (disjoint) automorphically-equivalent classes of solutions. A physical model corresponds then to a point in the orbit space $\mathcal{M}(S)/Perm(S)$ of all automorphically-equivalent classes of sections on $S$ . The projection map $\Pi:\mathcal{M}(S)\longrightarrow \mathcal{M}(S)/Perm(S)$ identifies all authomorphically-equivalent models to a single authomorphically-equivalence class. The image $\Pi(\sigma)=[\sigma]$represents the physical model defined by the model $\sigma$.

\section{Individuation Problem}

The hole argument for covariant and permutable theories is intimately related with the individuation problem of the base elements ( e.g., spacetime points or elementary particles). This is a consequence of the fact that the automorphisms of the base space provides the gauge freedom of the fields.

Einstein found a way to evade the hole argument, which translated in the language of manifolds says that one must assume that, at least inside the hole, the points of the manifold cannot be individuated independently of the metric field. This means that when we pull back (or drag along) the metric, we actually drag along the physically individuating properties and relations of the spacetime points. This is the reason why the pull back metric must not differ physically from the original metric.

The same reasoning applies to covariant and permutable theories: the base points may be characterized as such independently of the particular relations in which they stand, but they are individuated in terms of the relational structure given by sections of some fibered manifold/set defines the theory. 

Therefore, elementary particles, as spacetime points, must also be individuated by their position in a relational structure. For example electrons as a kind may be characterized in a way that is independent of the relational structure in which they are imbricated by their mass, spin and charge, for example; but a particular electron can only be individuated by its role in such a structure. \footnote{One should require, of course, that all relations between these particles are invariant under the permutation group acting on them.}

\end{document}